\documentclass[10pt,conference]{IEEEtran}
\IEEEoverridecommandlockouts
\usepackage{tabularx}
\usepackage{stfloats}
\usepackage[most]{tcolorbox}
\usepackage{float}
\usepackage{xspace}
\usepackage{subcaption}
\usepackage{graphicx}
\usepackage{courier}
\usepackage[ruled,vlined]{algorithm2e}
\usepackage{graphicx}
\usepackage{paralist}
\usepackage[shortlabels]{enumitem}
\usepackage{balance}
\usepackage{multirow}
\usepackage{multicol}
\usepackage{pbox}
\usepackage{mathtools}
\usepackage{algorithmic}
\usepackage{paralist}
\usepackage{xcolor}
\usepackage{url}
\usepackage{multirow}
\usepackage{multicol}
\usepackage{amsmath}
\usepackage{cite}
\usepackage{setspace}
\usepackage{flushend}
\usepackage{enumitem}
\usepackage{verbatim}
\usepackage{float}
\usepackage[normalem]{ulem}
\usepackage{xspace}
\usepackage{ifthen}
\usepackage{hyperref}
\usepackage{pifont}
\usepackage{tikz}
\usepackage{enumitem}
\usepackage{array}
\usepackage{multirow}

\usepackage{booktabs}
\usepackage{amsmath,amssymb}
\newcommand{\ie}{i.e.,}

\newcommand{\etal}{et al.}  

\newcommand{\OurComment}[1]{}
\newcommand{\Space}[1]{}

\newcommand{\Num}[1]{#1}

\newcommand{\posttabcaptionspace}{\vspace{-1ex}}

\newcommand{\accuweather}{AccuWeather}
\newcommand{\autoscout}{AutoScout24}
\newcommand{\duolingo}{Duolingo}
\newcommand{\flipboard}{Flipboard}
\newcommand{\merriamwebster}{Merriam-Webster}
\newcommand{\spotify}{Spotify}
\newcommand{\tripadvisor}{TripAdvisor}
\newcommand{\trivago}{trivago}
\newcommand{\wattpad}{Wattpad}
\newcommand{\webtoon}{WEBTOON}
\newcommand{\wish}{wish}
\newcommand{\youtube}{YouTube}

\newcommand{\classifierAccuracy}{81.4\%\xspace} 
\newcommand{\percImprovAPE}{11.0\%\xspace} 
\newcommand{\percImprovVET}{19.6\%\xspace}
\newcommand{\heurPassRate}{88.8\%\xspace}

\newcommand{\apeStopAverage}{275.1\xspace}

\def\z{\phantom{0}}

\newcommand{\kayak}{KAYAK\xspace}
\newcommand{\walmart}{Walmart}
\newcommand{\alltrails}{AllTrails}
\newcommand{\foxnews}{Fox News}
\newcommand{\carmax}{CarMax}

\newboolean{showcomments}

\setboolean{showcomments}{true}

\ifthenelse{\boolean{showcomments}}
  {\newcommand{\nb}[2]{
    \fbox{\bfseries\sffamily\scriptsize#1}
    {\sf\small$\blacktriangleright$\textit{#2}$\blacktriangleleft$}
   }
   
  }
  {\newcommand{\nb}[2]{}
   
  }

\newcommand{\safwat}[1]{{\color{black}{#1}}}
\newcommand{\safwatcrc}[1]{{{#1}}}

\newcommand\revision[1]{{{#1}}}

\newcommand{\aurora}{{{\sc Aurora}}\xspace}
\newcommand{\AIGtools}{AIG tools}
\newcommand{\AIGtool}{AIG tool}

\newcommand\MyBox[2]{
  \fbox{\lower0.75cm
    \vbox to 1.7cm{\vfil
      \hbox to 1.7cm{\hfil\parbox{1.4cm}{#1\\#2}\hfil}
      \vfil}
  }
}

\begin{document}

\title{A{\LARGE URORA}: Navigating UI Tarpits via\\Automated Neural Screen Understanding\thanks{This work was supported in part by Dragon Testing Technology and NSF grants CCF-1955853 and CCF-2132285.
Jiangfan Shi is also with the Institute of Computing Innovation (Zhejiang University),
Hangzhou, China.}}

\author{\IEEEauthorblockN{Safwat Ali Khan}
\IEEEauthorblockA{\textit{George Mason University}\\
Fairfax, VA, United States \\
skhan89@gmu.edu}
\and
\IEEEauthorblockN{Wenyu Wang}
\IEEEauthorblockA{\textit{University of Illinois}\\
Urbana, IL, United States \\
wenyu2@illinois.edu}
\and
\IEEEauthorblockN{Yiran Ren}
\IEEEauthorblockA{\textit{Dragon Testing Technology} \\
Hangzhou, China \\
renyiran@dragontesting.cn}
\and
\IEEEauthorblockN{Bin Zhu}
\IEEEauthorblockA{\textit{Dragon Testing Technology} \\
Hangzhou, China \\
zhubin@dragontesting.cn}
\and
\IEEEauthorblockN{Jiangfan Shi}
\IEEEauthorblockA{\textit{Dragon Testing Technology} \\
Hangzhou, China \\
shijiangfan@dragontesting.cn}
\and
\IEEEauthorblockN{Alyssa McGowan}
\IEEEauthorblockA{\textit{Thomas Jefferson High School} \\
Alexandria, VA, United States \\
alyssamc38@gmail.com}
\and
\IEEEauthorblockN{Wing Lam}
\IEEEauthorblockA{\textit{George Mason University} \\
Fairfax, VA, United States \\
winglam@gmu.edu}
\and
\IEEEauthorblockN{Kevin Moran}
\IEEEauthorblockA{\textit{University of Central Florida} \\
Orlando, FL, United States \\
kpmoran@ucf.edu}}

\maketitle

\pagestyle{plain}
\thispagestyle{plain}

\begin{abstract}
Nearly a decade of research in software engineering has focused on automating mobile app testing to help engineers in overcoming the unique challenges associated with the software platform.
Much of this work has come in the form of Automated Input Generation tools (\AIGtools) that dynamically explore app screens. 
However, such tools have repeatedly been demonstrated to achieve lower-than-expected code coverage -- particularly on sophisticated proprietary apps.
Prior work has illustrated that a primary\Space{ underlying} cause of these coverage deficiencies is related to so-called \emph{tarpits}, or complex screens that are difficult to navigate. 

In this paper, we take a critical step toward enabling \AIGtools~to effectively navigate tarpits during app exploration through a new form of automated semantic screen understanding. That is, we introduce \aurora, a technique that learns from the visual and textual patterns that exist in mobile app UIs to automatically detect common screen designs and navigate them accordingly. The key idea of \aurora is that there are a finite number of mobile app screen designs, albeit with subtle variations, such that the general patterns of different categories of UI designs can be \textit{learned}. As such, \aurora employs a multi-modal, neural screen classifier that is able to recognize the most common types of UI screen designs. After recognizing a given screen, it then applies a set of flexible and generalizable heuristics to properly navigate the screen. We evaluated \aurora both on a set of 12 apps with known tarpits from prior work, and on a new set of five of the most popular apps from the Google Play store. 
Our results indicate that \aurora is able to effectively navigate tarpit screens, outperforming prior approaches that \textit{avoid} tarpits by 19.6\% in terms of method coverage. 
Our analysis of the results finds that the improvements can be attributed to \aurora's UI design classification and heuristic navigation techniques.
\end{abstract}

\vspace{-0.5em}
\section{Introduction}
\label{sec:intro}

Mobile application development (or app development) is a challenging endeavor. Developers working in this domain face a variety of unique challenges that range from rapidly evolving and fault-prone APIs~\cite{mario:fse13,Bavota:TSE15} to frequent user feedback~\cite{palomba:icsme15} and highly competitive app marketplaces~\cite{appbrain_android_2023}. As such, software maintenance and testing techniques, which are critical to ensuring software quality, are often overlooked due to pressure to deliver features in the face of these external factors~\cite{kochar:ICST15}. As such, the research community has worked to provide a range of automated techniques to help developers cope with these challenges, spanning tools that support tasks from bug management~\cite{chaparro2019assessing,Mahmud:ICSE'24,Yang:ICSE'23,Yan:ICSE'24,Yang:FSE'22,Fazzini:TSE22,Bernal:TSE23,Johnson:SANER22,Cooper:ICSE21,Havranek:ICSE21,Bernal:ICSE20} to software evolution~\cite{moran2018detecting,Salma:MSR24}.

In the past decade, an extremely popular area of work in the software engineering research community has aimed to automate \textit{software testing} for mobile apps -- more specifically, \textit{GUI-based testing}. Given that mobile apps are GUI-centric in nature, UI testing is one of the most popular testing modalities for ensuring the correctness of functionality. However, creating UI tests manually is extremely time-consuming~\cite{kochar:ICST15}. As such, the research community has developed numerous Automated Input Generation (AIG) tools that dynamically explore applications with the goals of exercising substantial portions of app functionality, while simultaneously uncovering crashes and other faults.\Space{
Based on the methodology of action planning, t}
These tools can be broadly grouped into \textit{random-based}~\cite{orso_software_2014,monkey_uiapplication_2022,gu_practical_2019,machiry_dynodroid_2013,sasnauskas_intent_2014,ravindranath_automatic_2014}, \textit{model-based}~\cite{moran_automatically_2016,machiry_dynodroid_2013,baek_automated_2016,monkey_uiapplication_2022,mao_sapienz_2016,dong_time-travel_2020}, and \textit{machine learning-based} tools~\cite{li_humanoid_2019,deka_rico_2017,zheng_automatic_2021}.

In controlled experimental settings, \AIGtools{} often perform well and achieve reasonably high code coverage. However, in practice, these \AIGtools{} are often prone to low effectiveness in certain testing scenarios, particularly on sophisticated proprietary apps~\cite{wang_empirical_2018}.
One reason for this low effectiveness is that many proprietary apps often contain complex screens that are difficult for AIG tools to navigate, i.e., the semantics of the screen require a precise order of actions to navigate and bypass so that additional app states can be explored.
Wang \etal~recently performed a study that empirically illustrated this phenomena, wherein they observed different types of screens that caused \AIGtools{} to halt exploration progress~\cite{wang_vet_2021}. 
The authors of this recent study refer to such screens as \textit{UI Exploration Tarpits}. In addition to demonstrating this phenomenon, the authors also introduced a preliminary technique for dealing with these tarpit screens, called VET.  
VET integrates with existing AIG tools and uses a learn-from-mistake strategy that first identifies exploration tarpits from runs of the \AIGtool{} and later disables 
these screens in future runs of the \AIGtool{}.
The VET approach introduced by Wang \etal{} has\Space{ at least} two major limitations. 
First, VET is inherently expensive to run in practice, as it requires running an AIG tool twice, once to detect potential tarpits and then again to explore the app with the tarpits disabled. 
Second, VET does not assist with \textit{navigating} UI tarpits, it simply disables them, meaning that there will always be portions of an application's state that AIG tools enhanced with VET cannot explore. 
However, navigating \textit{through} UI tarpits may be feasible (as suggested by the results of the manual analysis performed by Wang \etal~\cite{wang_vet_2021}), exploration tarpits often fall into one of a limited number of categories, even across different apps and \AIGtools{}. 
This finding suggests that there may be patterns that can be exploited to explore these complex UI tarpit screens.

Given the findings and current limitations of prior work, in this paper, we propose a novel technique called \aurora, that aims to effectively \textit{navigate} tarpit screens during app exploration using a new form of automated semantic screen understanding. 
The key idea of \aurora is that there are a finite number of UI designs for screens that are likely to represent tarpits, and hence, general \textit{design motifs} that can be learned will allow for the automated recognition and navigation of such screens. 
\aurora learns from both the visual and textual patterns present in UI screens to automatically identify tarpit screens, through a component we call the \textit{screen recognizer}, and then navigates recognized screens using a set of flexible heuristics, via the \textit{heuristic navigator} component.

To better understand UI design categories and their relationship with exploration tarpits, we first studied Android app screenshots and UI hierarchies from the RICO dataset~\cite{deka_rico_2017}, deriving a set of \Num{21} mobile app \textit{UI Design Motifs}, representing coherent design patterns. During this process, multiple authors jointly labeled a minimum of 60 screens exhibiting each of our 21 design motifs, for a dataset totaling 1369 UI screens.
We then proceeded to study the correlation between these general categories of UI designs and the tarpit screens as manifested through the dataset of tarpits discovered by and published alongside the VET tool~\cite{wang_vet_2021}. We found that eight of our 21 design motifs were identified as tarpits in the VET dataset. 

\newcommand{\screenrecog}{screen recognizer\xspace}

This analysis of the relationship between various categories of UI designs and tarpits inspired our design of \aurora. 
During the app exploration process of an AIG tool, \aurora is able to detect when app exploration progress is hindered by a tarpit, and will then automatically categorize the tarpit screen into one of our identified categories using a neural screen understanding approach that analyzes \textit{both} the visual patterns in a screenshot of the tarpit and the textual patterns on the UI.
For example, \aurora's \screenrecog can categorize a given screen as a login screen if the screenshot and UI\Space{ properties} contain a username and a password \texttt{\small\textbf{EditText}}, and contain salient visual patterns that indicate the typical structure of a login screen, such as center-aligned text-box(es) and button(s).

\safwatcrc{\aurora's \screenrecog is implemented using a multi-modal deep learning model~\cite{radford_learning_2021}, which is initialized through self-supervised learning on a set of \Num{6000} app screenshots extracted from an online search engine. It is then trained and tested on an 80-20 split of our labeled dataset of 1369 UI images. Our evaluation finds that \aurora achieved \Num{\classifierAccuracy} classification accuracy.}
\aurora's \textit{heuristic navigator} implements \Num{eight} input generation heuristics that aim to intelligently generate sequenced input scenarios to navigate through typical tarpit categories using a combination of neural text matching via transformer-based language models, and\Space{ automated} dynamic analysis.

\aurora is designed to run alongside and enhance existing \AIGtools{} by quickly identifying and navigating through identified exploration tarpits.
As an \AIGtool{} is exploring, \aurora will periodically check if the exploration is stuck.
If the \AIGtool{} appears to be stuck, \aurora will analyze the current screen to determine if it is a tarpit.
If it is, then \aurora will pause the \AIGtool{} and activate a corresponding input generation heuristic to navigate the current screen.
We combine \aurora with a state-of-the-art \AIGtool{}, APE~\cite{gu_practical_2019}, and test on 17 popular Android apps.
We find that \aurora helps improve code coverage substantially, with an average improvement of \percImprovAPE over APE and \percImprovVET over VET \safwatcrc{on two separate comparative analyses}. Through a qualitative analysis, we observe that these improvements arise due to the effectiveness of \aurora's \screenrecog and heuristic navigation strategies -- the latter of which exhibits an \heurPassRate success rate in navigating through UI tarpit screens.

\aurora was developed in cooperation with Hangzhou Dragon Testing Technology Co., Ltd. who is focused on building AI-enhanced software testing products, and customers of the company include internationally recognized clients, such as WeChat, a messenger app with over one billion monthly active users. Dragon Testing has deployed a proprietary version of \aurora, that closely mirrors the components and workflow described in this paper, to its automated software testing product offerings. In this context, \aurora has enabled automated navigation of several types of tarpit screens for thousands of automated test cases, including pop-ups and forms, that had previously hindered the testing of app business logic for Dragon Testing's customers. This deployment of \aurora further illustrates both its effectiveness and practicality.

In summary, this paper makes the following contributions:
\begin{itemize}
\item \textbf{A Study} identifying prevalent \textit{design motifs} of Android UI screens and the categories that constitute UI exploration tarpits.
\item \textbf{A Multi-modal Deep Learning-based Approach} for classifying a given UI screen into the design motifs identified in our study.
\item \textbf{Automated Heuristics} that can be used to navigate prevalent UI tarpits.
\item \textbf{A Framework} implemented as \aurora~\cite{appendix}, which can be combined with automated input generation (AIG) tools to automatically categorize screens and apply relevant heuristics to help AIG tools bypass UI tarpits.
\item \textbf{An Evaluation} on the (1)~accurateness of our deep learning model at predicting UI screen categories, (2)~effectiveness of our heuristics at bypassing UI tarpits, and (3)~code coverage improvements that \aurora can help AIG tools achieve.
\item \textbf{Artifacts} made publicly available from this work, which includes \aurora and our labeled dataset of UI categories, to help aid future research\Space{ researchers for easy replication, testing real-world Android apps, or building upon for future research.}~\cite{appendix}.
\end{itemize}
\section{Background}
\label{sec:background}

 \begin{figure}[t]
  \centering
  \includegraphics[width=0.85\linewidth]{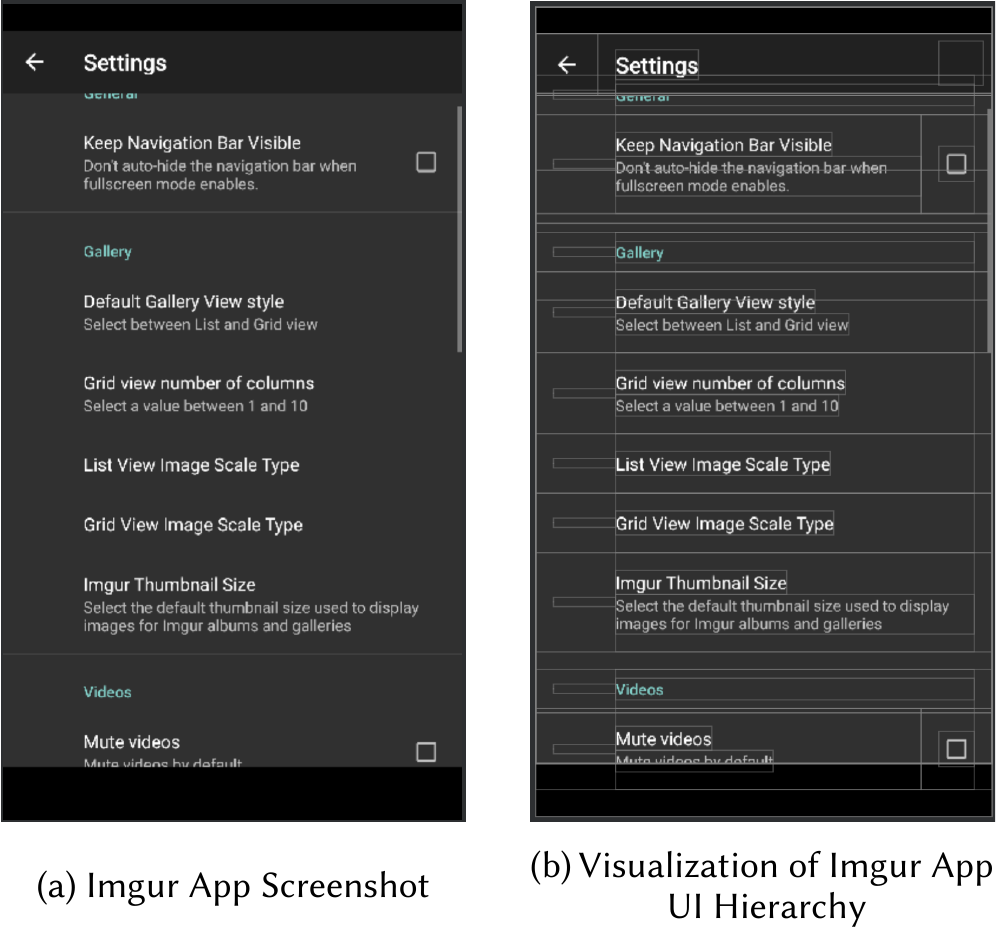}
  \vspace{-0.5em}
  \caption{A visualization of the spatial properties of components captured within Android UI hierarchy of the Imgur\cite{imgur} app.}
  \label{fig:ui-hierarchy}
\end{figure}

\subsection{Mobile App UI Hierarchies and Frameworks}
A \textit{UI Hierarchy} represents the contents of an app UI rendered to the screen of a mobile device.
UI hierarchies are comprised of UI elements that each exhibit several properties (e.g., location, size, component type), wherein UI elements are arranged in a hierarchical fashion and children can inherit design and logical properties from their parents.
In Android, the \texttt{\small\textbf{ViewServer}} generates and maintains runtime information about app UIs, and can be queried using the \texttt{\small\textbf{uiautomator}} framework, which extracts a representation of the UI hierarchy in \texttt{\small\textbf{xml}} format. Figure~\ref{fig:ui-hierarchy} illustrates a visualization of the spatial properties of components captured as output in \texttt{\small\textbf{uiautomator}} \texttt{\small\textbf{xml}} files. Programmatically, Android screens are primarily made up of constructs called \textit{Activities} and \textit{Fragments}, where Activities typically represent a single logical screen and Fragments represent smaller components of a screen, such as a pop-up menu. For the purposes of discussion in this paper, when we refer to a ``UI screen'', we are effectively referring to a single Activity or Fragment that is rendered to the UI screen, such as the ``Settings Activity'' illustrated in Figure~\ref{fig:ui-hierarchy}.

UI hierarchies effectively capture information that is relevant to the \textit{structure} of a UI screen, which we will illustrate is important for learning abstract representations of UIs to aid in \aurora's screen classification capabilities. In addition to UI hierarchies, screenshots can be easily captured from mobile apps using the \texttt{\small\textbf{screencap}} utility built into the Android Debug bridge (\texttt{\small\textbf{adb}}) framework. 
This captured UI information is often used by \AIGtools{} for decision-making and is especially critical to model-based testing tools. As we describe later, \aurora extracts runtime UI hierarchies and screenshots via the \texttt{\small\textbf{uiautomator}} and \texttt{\small\textbf{adb}} frameworks, which are fed as input to both \aurora's \textit{screen recognizer} and \textit{heuristic navigator}.

\subsection{UI Exploration Tarpits and the VET Approach}
\begin{figure}[t]
  \centering
  \includegraphics[width=0.81\linewidth]{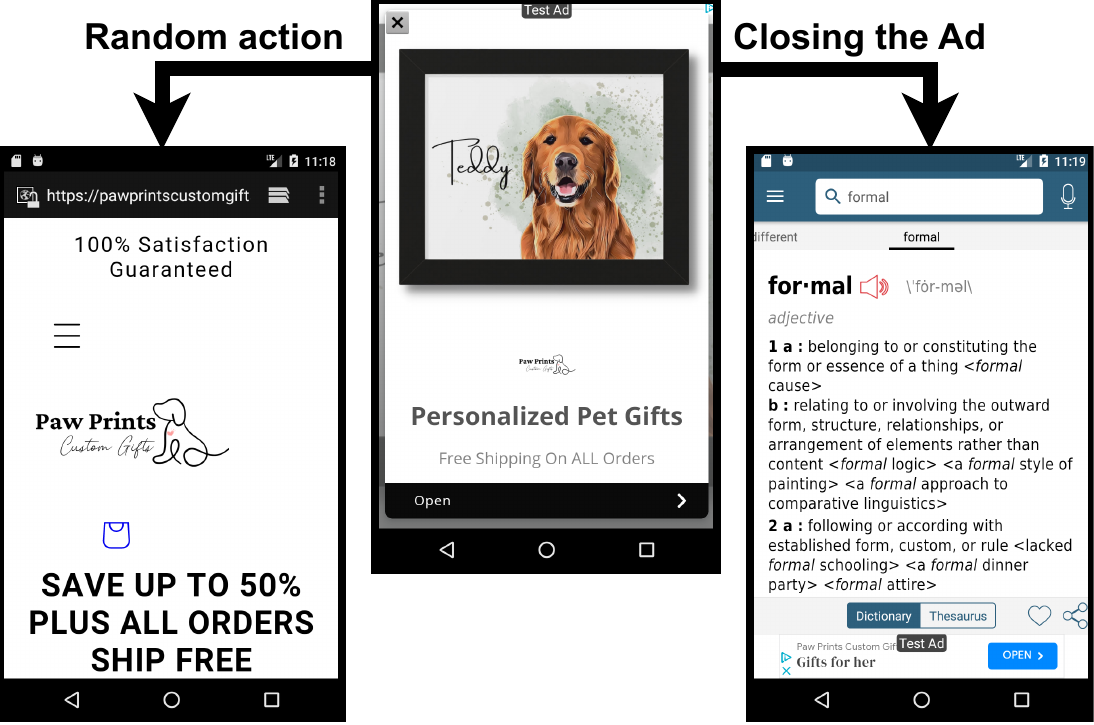}
  \vspace{0em}
  \caption{Example of an Advertisement screen tarpit.}
  \label{ad_screen}
\end{figure}

{\it UI Exploration Tarpits} (or UI Tarpits) is a term coined by Wang \etal~\cite{wang_vet_2021}, that describes a phenomenon that occurs when an \AIGtool{} explores a single UI screen (\ie~\textit{Activity}) for an excessive amount of time, hindering the progress of the tool in exploring undiscovered UI states of a given app. 
\safwatcrc{Figure~\ref{ad_screen} shows an example advertisement tarpit screen in the \merriamwebster~app. Random actions on these screens typically cause an AIG tool to get stuck -- in this example, the advertisement launches a WebView that can only be closed through a specific series of actions.}

To overcome issues with these screens, Wang \etal~proposed the VET tool~\cite{wang_vet_2021} to detect and cope with UI exploration tarpits.
VET works in three stages:
First, it runs the \AIGtool{} on the target app without any restriction and records the testing process, yielding traces that record how the \AIGtool{} has interacted with the target app.
Second, it analyzes the collected traces against pre-defined low-level patterns that characterize repetitions to discover potential exploration tarpits.
A ranking strategy is utilized to reduce false positives.
Third, it reruns the \AIGtool{} on the target app with the discovered exploration tarpits disabled, achieved through dynamically blocking the entry points to the exploration tarpits or via restarting the target app.
Evaluation results show that VET can improve the testing effectiveness of multiple Android \AIGtools{}, and can help improve bug detection capabilities. 

Despite its advancements, VET has two major limiting factors. First, VET must first run an AIG tool \textit{comprehensively} to identify and build a model of potential UI tarpits. This requirement means that testing time for apps is often \textit{doubled} due to the need to run once to detect tarpits, and another time to explore the app with tarpits disabled. Second, VET does not facilitate \textit{navigating} through tarpit screens, but instead, simply \textit{disables} tarpit screens. 
This limitation means that VET is effectively incapable of exploring certain areas of an app that were identified as tarpits. \aurora aims to overcome both of these shortcomings by analyzing UI screens in real-time to detect, classify, and navigate through tarpit screens using new forms of automated neural UI screen understanding.

\section{Deriving Mobile App UI Design Motifs}

One of the key ideas underlying \aurora is that certain designs for mobile app UI screens are \textit{reused}, with varying degrees of change and variability, across applications such that the structural and lexical patterns of these designs can be automatically identified. 
In this paper, we refer to these patterns as \textit{UI Design Motifs}. 
To better understand the types and prevalence of UI Design Motifs across Android applications, we conducted a preliminary investigation to derive a set of motifs that will serve as the focus of our \aurora approach.

\subsection{Isolating Structural Screen Patterns with Silhouette Screens}
\label{subsec:silhouette}

To empirically derive a set of UI design motifs, we analyzed a set of randomly sampled screens from the RICO dataset~\cite{deka_rico_2017}. RICO is currently the largest dataset of Android app UI screenshots and corresponding runtime UI metadata, spanning over 66,000 UI screens collected from over 9,000 free Android apps available on the Google Play Store. 
This screen information was collected via a combination of automated UI exploration and crowdsourced UI exploration on virtual Android devices. The UI metadata for these screens was collected using the \texttt{\small\textbf{uiautomator}} framework. 

Given the sheer scale of this dataset, manually analyzing even small portions of the dataset to discover UI design motifs would be a time-consuming proposition. Furthermore, as observed by the authors of the RICO dataset, there likely exist certain common UI patterns followed by a long tail of unique or one-off UI screen designs. Given that our aim is to identify and categorize common UI design patterns into categories we term as \textit{motifs}, we introduced lightweight automation to facilitate the manual labeling process. As such, we conducted an initial step of unsupervised, computer vision-based clustering of screens into broad categories that exhibited visual similarities.

 \begin{figure}[t]
  \centering
  \includegraphics[width=0.65\linewidth]{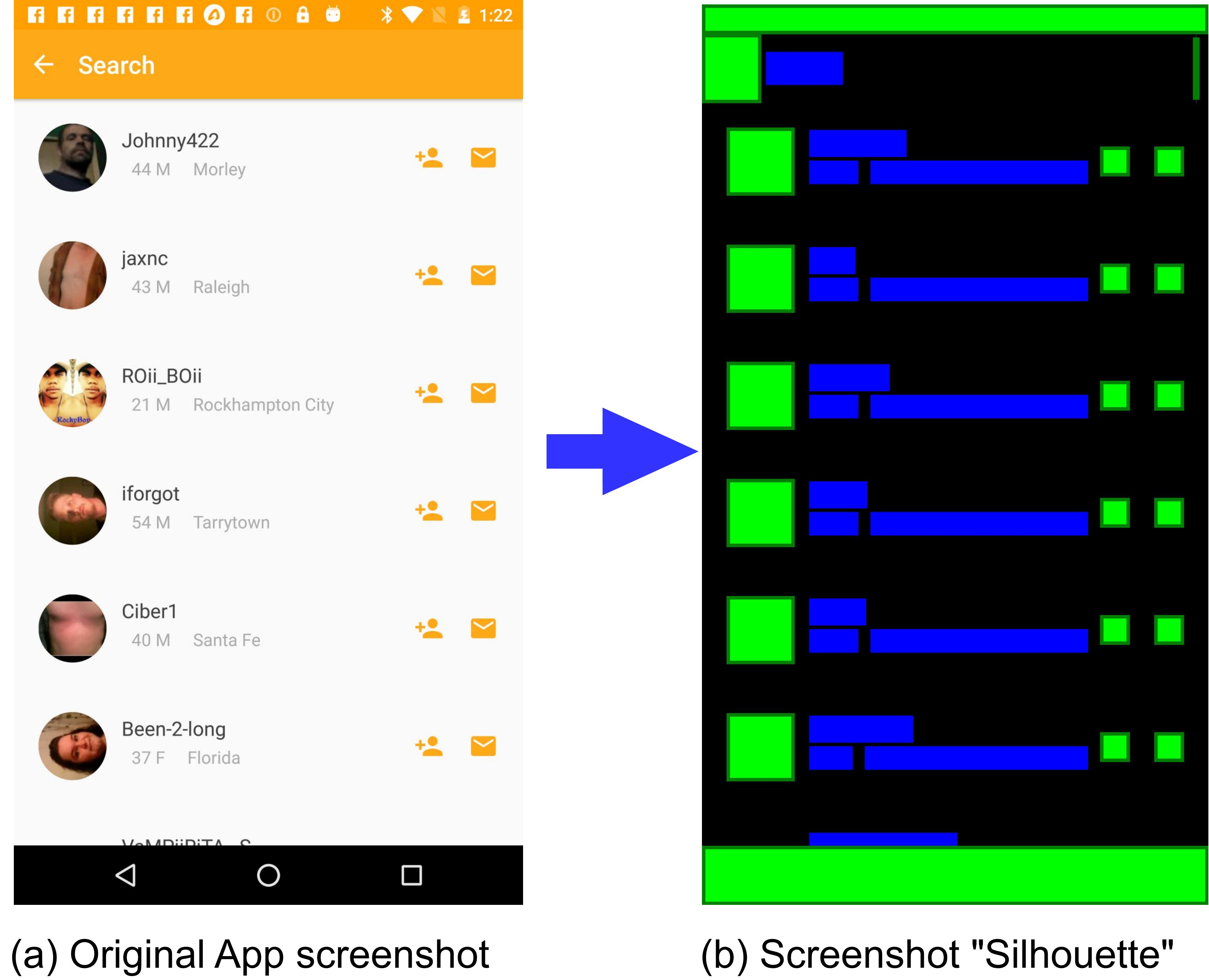}
  \caption{A visualization of a UI Silhouette Screen.}
  \label{fig:silhouette}
\end{figure}

Using ``raw'' screenshots to group together visually similar UI screens for the purposes of deriving design motifs presents certain challenges. 
First, two screens that share a design motif (\ie~Settings Screen) may \textit{instantiate} that screen using a similar screen \textit{structure}, but have widely varying colors, fonts, and other stylistic properties. 
As such, clustering screens according to raw image similarities is likely to be greatly impacted by similarities in style. To focus our analysis on \textit{structural} UI Design Motifs that occur across a diverse set of apps, we developed a process to ``abstract'' raw UI screenshots, by stripping out stylistic information and creating what we refer to as \textit{Silhouette Screens}. Figure~\ref{fig:silhouette} illustrates this process. In essence, we divide all UI components into one of two categories: (i) textual components, and (ii) non-textual components. \safwatcrc{We then draw textual components on a black canvas as blue boxes and non-textual components as green boxes, }
parsing the spatial component size and locations from RICO's \texttt{\small\textbf{uiautomator}} metadata that accompanies each screenshot. This methodology allows us to effectively capture each screen's abstract structure, while ignoring the stylistic variations different screens may exhibit. We discuss different potential variations of Silhouette Screen creation in Section~\ref{sec:approach}.

\subsection{Structural Clustering of UI Screens}
\label{subsec:screen-clustering}

After developing our process for creating Silhouette Screens, we needed a reliable methodology to create a robust representation of our UI screens so that they could be clustered according to their structural UI features. 
While simple image similarity measures could be used to accomplish this goal, such measures often rely on handcrafted heuristics that may not transfer well to UI screens. Furthermore, deep learning-based computer vision techniques that incorporate convolutional neural networks have been shown adept at learning robust representations of image features~\cite{alexnet,He:ResNet}. As such, we implemented a 2D convolutional autoencoder, consisting of 6 encoder and 7 decoder layers, and trained it on 32,338 UI screens collected by randomly sampling from 50\% of the  RICO dataset.
Generally speaking, an autoencoder is a neural network that encodes an image into a high-dimensional vector representation and then is trained to decode this representation back into the original image, with differences between the input and output being used by a loss function to update model weights during the training process. We trained our autoencoder model on a subset of RICO to avoid overfitting and to improve the model's generalization to handle a wider range of real-world data. Through experimentation, we also found that this amount of data was sufficient to train our autoencoder to convergence. 

We then applied a K-means clustering technique to 645 screens, roughly representing $\approx$1\% of the RICO dataset sampled from outside the autoencoder's training set. Using the elbow technique~\cite{elbow}, we found 30 clusters to be optimal.

\subsection{Manual Analysis and Derivation of Design Motifs}
\label{subsec:screen-clustering-motifs}

After this clustering procedure, three authors of this paper manually analyzed each of the 30 clusters, refined the categorizations of screens, and then collectively provided labels to the finalized clusters. This process proceeded as follows: First, separate authors would look at a defined set of clusters (\ie~five clusters), and they would examine each screen in these clusters and re-cluster them\Space{ by moving images} or form new clusters to better group screens that share a common structural UI pattern. Two authors would examine these sets of clusters, and then all three authors would meet to discuss the results and come to an agreement on the newly formed clusters. This process proceeded until all 30 clusters had been examined. The end result of this process was a refined set of 21 clusters, representing UI \textit{design motifs} with the following labels: \textit{Advertisement}, \textit{Calendar time and weather}, \textit{Catalog}, \textit{Feed}, \textit{Form}, \textit{Home menu}, \textit{List}, \textit{Log-in}, \textit{Map}, \textit{Onboarding}, \textit{Player}, \textit{Pop up}, \textit{Product}, \textit{Search}, \textit{Settings}, \textit{Splash}, \textit{Terms and conditions}, \textit{Travel booking}, \textit{Type message}, \textit{Viewer}, and \textit{Web browser} screens. We provide complete descriptions and examples of these categorizations in our artifact~\cite{appendix}.

 \begin{figure}[t]
  \centering
  \includegraphics[width=0.8\linewidth]{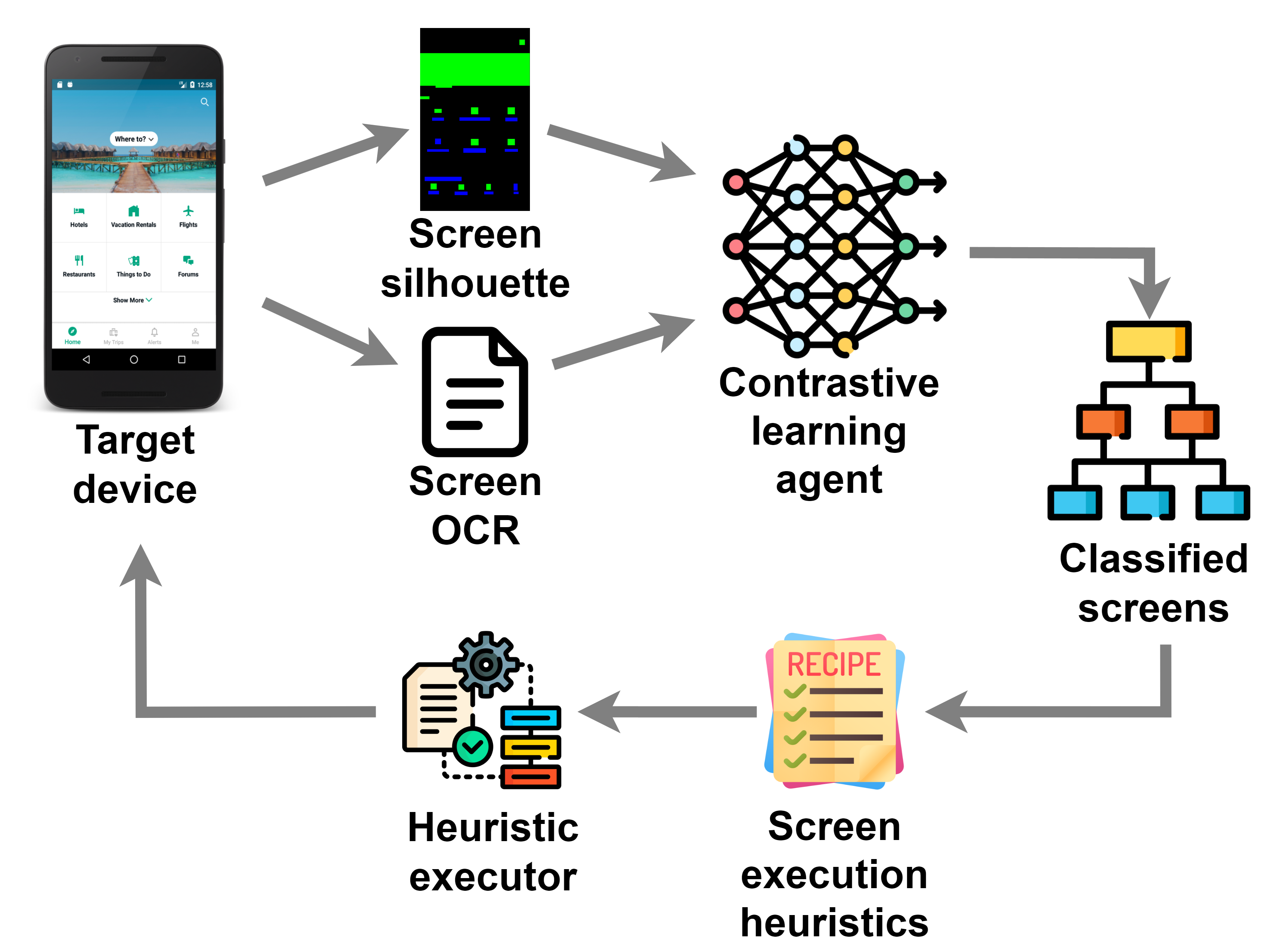}
  \vspace{-1ex}
  \caption{Overview of \aurora's workflow.}
  \vspace{1ex}
  \label{auroraOverview}
\end{figure}

\subsection{Identifying Tarpit Screens and Motifs from the VET Dataset}
\label{subsec:vetdataset}

The study conducted by Wang et al.~\cite{wang_vet_2021} contains a dataset of over 95,000 UI screens obtained from testing 16 different industrial apps using different AIG tools, including APE~\cite{gu_practical_2019}, Monkey~\cite{monkey_uiapplication_2022}, and WCTester~\cite{zheng_automated_2017}. These UI screens are organized into 127 test traces, with each trace comprising a collection of screenshots and associated metadata derived from a specific app-tool pairing. These traces include timing information as well as the number of actions executed on various UI screens, allowing us to identify screens for which the tools encounter obstacles that may represent tarpits.

To find examples of tarpit UI screens, we perform a UI metadata comparison using the {\texttt{\textbf{\small JSONComparison API}}~\cite{karpushkin_jsoncomparison_2023} on the \texttt{\small\textbf{uiautomator}} metadata included with each screenshot from the VET traces. We set a minimum threshold  of five action repetitions \textit{and} 10 seconds of elapsed time to consider a screen as a tarpit in the context of the traces. Said differently, we only consider a screen to be a tarpit if at least five consecutive actions were entered on the screen and the screen did not change for at least 10 seconds. Applying these criteria to the screens from the traces, we extract 238 screens that likely represent UI tarpits. Upon further inspection of this dataset, we observed that there were screens with fewer actions but very large elapsed time. Therefore, we additionally include the top 200 screens on which the tools used in VET's evaluation spent the most time. In the end, we take the union of these two sets which in total contains 348 tarpit screens.

\subsection{Tarpit Screen Analysis}
\label{analyzeVET}

We aimed to better understand two aspects of the tarpit screens from the VET dataset including: (i) reasons for obstructed AIG tool navigation, and (ii) the labels of these screens according to the design motifs we derived earlier. To carry out this process, two authors of the paper independently analyzed each of the 348 tarpit screens and investigated the two factors above. We present these findings below.

\begin{tcolorbox}
\footnotesize
\vspace{-0.5em}
\textbf{Reasons for AIG Tool Obstructions:}
\begin{enumerate}[leftmargin=*]
    \item Inability to input relevant text in a designated field.
    \item Difficulty locating specific components or series of components for screen progression.
    \item Failure to identify interactive elements on the screen.
    \item Accidental engagement with advertisements, resulting in difficulty exiting the web view screen.
    \item Tool initiated closure of the application.
    \item App unresponsiveness.
\end{enumerate}
\vspace{-0.7em}
\end{tcolorbox}

\begin{tcolorbox}
\footnotesize
\vspace{-0.5em}
\textbf{Tarpit Screen Design Motifs and Navigation Strategies:}
\begin{itemize}[leftmargin=*]
    \item \textbf{Log-in:} Input some predefined text and tap certain buttons.
    \item \textbf{Onboarding:} Identify and interact with certain buttons.
    \item \textbf{Player:} Perform sequential actions to access different screens.
    \item \textbf{Advertisement:} Locate and tap the close button to exit.
    \item \textbf{Viewer:} Tap on the screen to expose interactive elements or use the back button for navigation.
    \item \textbf{Form:} Input relevant text, interact with spinners and buttons.
    \item \textbf{Web browser:} Tap the back button or restart the app.
    \item \textbf{Search:} Input relevant text in search fields.
\end{itemize}
\vspace{-0.5em}
\end{tcolorbox}

\section{The Aurora Approach}
\label{sec:approach}

In this section, we define the methodology of our Android UI exploration tool, \aurora ~\cite{appendix}. 
Our approach leverages a multi-modal computer vision-based \textit{screen recognizer} that uses structural and lexical patterns in UI screens to detect different UI Design Motifs. When an AIG Tool encounters a tarpit screen, \aurora classifies this screen using the screen recognizer and then applies one of several pre-defined \textit{navigation heuristics} to navigate through the screen to uncover additional states of the application. An overview of the approach is shown in Figure~\ref{auroraOverview}. We describe the components of \aurora in detail in the following subsections. 

\subsection{Approach Workflow}

\aurora \cite{appendix} operates in conjunction with any AIG Tool that extracts \texttt{\small\textbf{uiautomator}} metadata and screenshots as part of its exploration process. The integrated AIG tool can be random-based, model-based, or machine learning-based in nature. 
During an AIG tool's app exploration, we need a mechanism to detect when the tool may be ``stuck'' in a tarpit screen. We derived a suitable elapsed time to determine whether a screen is a tarpit empirically through a small set of experiments where we varied the ``tarpit trigger'' time from 10 seconds to 30 seconds in 5-second intervals, and inspected the number of identified tarpits during a one-hour execution of the APE~\cite{gu_practical_2019} automated testing tool on the set of 16 apps from the VET dataset. Upon inspection of the identified screens at each tarpit threshold level, we found that 10 seconds allowed \aurora to detect the highest number of true tarpit screens with a reasonably small number of false positives. As such, \aurora polls the activity/window combination queried from an Android emulator’s view-server every second –- if the activity/window combination remains the same after 10 seconds \aurora is triggered to bypass the potential tarpit.

Once a tarpit screen has been identified, a screenshot and the corresponding \texttt{\small\textbf{uiautomator}} metadata for the screen are saved, and then a Silhouette Screen is created to capture the structural patterns, and the EAST Optical Character Recognition (OCR) technique~\cite{EAST} from the Google Cloud Vision API~\cite{google-ocr} is used to extract text from the screen. The \texttt{\small\textbf{uiautomator}} metadata, Silhouette Screen, and OCR data are then converted to visual and textual embeddings and passed to \aurora's \textit{screen recognizer}. This component generates a ranked list of UI Design Motifs for the current screen.

Given the ranked set of potential UI Design Motifs for a given screen, \aurora checks if the screen falls into one of the eight Tarpit UI Design Motifs derived in Section~\ref{analyzeVET}. 
If so, then \aurora triggers the execution of a \textit{navigation heuristic} for that Design Motif, which performs a predefined set of UI actions. 
To carry out the actions of a heuristic, \aurora uses the SentenceBERT model~\cite{reimers_sentence-bert_2019} to determine where to input predefined text and where to click.
It is possible that a given tarpit screen could be mis-classified or that a given heuristic might fail. 
In these scenarios, if \aurora recognizes that it has not navigated through a given UI screen after the execution of a heuristic, it then executes heuristics of the next \textit{two} tarpit categories from the ranked list of predicted UI Design Motifs. 
Given that \aurora's heuristics generate fewer actions per given unit of time than many AIG Tools, it is necessary to limit the number of heuristics executed on a given tarpit screen to small, reasonable number (three)\Space{ in the worst-case scenario where the first two heuristics do not advance the UI exploration process}. 
If no progress is made after three heuristic execution attempts, the app is restarted.

\subsection{UI Screen Recognizer}
\label{approach:classifer}

As part of our development of \aurora, we worked with Dragon Testing to develop two different UI screen classifiers. The first classifier combines image embeddings learned from a neural autoencoder, with lexical embeddings of screen text from a large language model. 
These embeddings are concatenated and then passed into a classifier that predicts a ranked list of screens. 
The second classifier uses a multi-modal CLiP~\cite{radford_learning_2021} model to encode images and text. 
We describe each of these classifiers below. 
We explore these two classifiers as they require very different numbers of parameters, with the CLiP-based approach requiring a much larger number of parameters than the Autoencoder-based approach. 
Given that past research has suggested the exploration of machine learning techniques of varying degrees of complexity when applied to software data~\cite{menzies-ml}, we opted to explore both a ``simpler'' and ``more advanced'' machine learning technique. 
We present the results of an empirical comparison of these two classifiers in Section~\ref{sec:eval}.

\subsubsection{Autoencoder-based UI Design Motif Classifier}

The first classifier that we constructed for \aurora uses both a visual and textual classifier before combining the output of these two techniques to produce a final ranked list of categorized UI Design motifs for a given target screen.

For the visual classifier, we utilize the encoder portion of our pre-trained autoencoder framework described in Section~\ref{subsec:screen-clustering}. The encoder transforms high-dimensional image data into compact feature vectors. These feature vectors contain essential information from the images while reducing their dimensionality. Leveraging these feature vectors, we applied traditional machine learning models, such as Random Forest, Multi-Layer Perceptron (MLP), and Naive Bayes to conduct classification tasks on a set of labeled UI screens. 
In our experiments, we found the Random-forest classifier to achieve the highest accuracy. 
To derive a training set for this approach, we had two authors label an additional 599 screenshots from the RICO dataset into our 12 Design Motif categories, distributed as evenly as possible.

For the textual classifier, we use both the UI metadata and the OCR output from the UI screenshots and create a single paragraph out of each UI screen. 
To encode the text from the UI metadata, we convert six different component attributes for each component in the UI hierarchy into sentences. 
The attributes that \aurora encodes are (i) component class, (ii) ancestor class, (iii) label, (iv) position, (v) width, and (vi) height.
In cases where \textit{labels}, which typically represent text rendered to the screen, are missing in the UI metadata, we utilize OCR information using Google's Tesseract OCR engine~\cite{noauthor_tesseract_2023} as an alternative source of data. We applied various classifiers, including Naive Bayes, KNN, Random Forest, and Multi-Layer Perceptron (MLP), in conjunction with a TF-IDF vectorizer. The TF-IDF vectorizer combined with the MLP classifier produced the highest accuracy. We utilized the same training set of 599 screenshots for this textual classifier\Space{ as well}.

Finally, \aurora combines the output of these two classifiers together. It uses a combined probabilistic approach to predict screen class by extracting 21 attributes from both the visual and textual classifiers, each denoting a probability value for a specific class. Subsequently, we utilize a Random Forest classifier for prediction.

\subsubsection{Transformer-based UI Design Motif Classifier}
\label{sec:approach:classifier}
\label{evaluation:classifer}

\begin{figure}[t]
  \centering
  \includegraphics[width=0.82\linewidth]{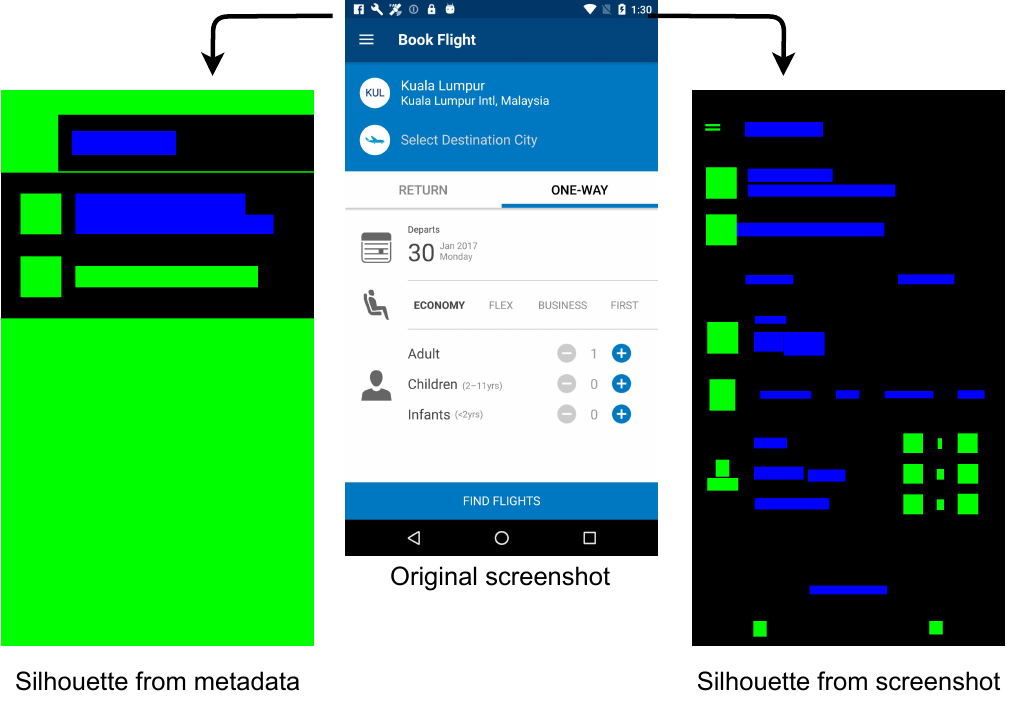}
  \caption{Comparing Silhouette Screens - blue color represents text and green color represents non-text components.}
  \label{fig:compSil}
\end{figure}

\aurora's transformer-based approach uses CLiP (Contrastive Language-Image Pre-Training)~\cite{radford_learning_2021}, a model known for its ability to perform transfer learning on a wide array of classification-based downstream tasks. 
The CLiP model works by creating two separate embeddings for the image and the provided text. It creates image embedding using a convolutional neural network (CNN) based on the ResNet architecture. For text embedding, it uses a transformer-based neural network architecture similar to the GPT language model \cite{radford_improving_2018}. Additionally, for this classifier, we improved upon the concept of Silhouette Screens we introduced earlier in the paper. That is, our prior technique for deriving Silhouette Screens relied \textit{solely} on \texttt{\small\textbf{uiautomator}} metadata to create the Silhouette Screen. However, as illustrated in Figure~\ref{fig:compSil}, there may be certain textual elements that are not properly captured in the UI metadata -- for instance if the text is displayed through an image or web view. As such, we use the Google Cloud OCR Engine to detect text on the screen, relate it to the spatial properties of components, and create more accurate UI Silhouette Screens, as shown in the right-hand side of Figure~\ref{fig:compSil}.

\safwatcrc{To train the CLiP model, we conduct unsupervised pre-training on a set of 6,000 UI screenshots that we crawl from Google. These screens were collected by searching the terms of each of our identified design motifs in conjunction with the terms ``mobile app screen'' using Google image search, and downloading the resulting UI screenshots until we had 6000 screens distributed across our design motif categories. We then train and test CLiP using an 80-20 split of our new set of 1369 RICO images (expanding from the set of 599 used to train the Autoencoder-based model) evenly distributed across our 21 Design Motif categories.}

To encode the textual information, we use the CliP model to extract textual embeddings from all the text displayed on a given UI screen as extracted by Google's Cloud OCR engine.

\subsection{Heuristics Design}
Our heuristics are formulated through the analysis of tarpit screens within the VET dataset. Considering the reasons for getting stuck at the end of Section~\ref{analyzeVET}, we created heuristics in executable Python code. 
The heuristic code makes use of the Python wrapper for Android Debug Bridge commands~\cite{ppadb} for sending commands to a virtual device on an Android emulator. 
While looking for specific components or input fields, it uses the SentenceBERT model~\cite{reimers_sentence-bert_2019} to match with the closest on-screen component. 
To validate the functionality and reliability of our executable heuristic code, we tested it rigorously on three Android applications, ensuring that they can execute without any operational issues or errors. 
\aurora focuses on eight specific types of UI screens, as detailed in Section~\ref{analyzeVET}, for which it has developed specific heuristics that are generalizable across various applications. We provide a description of two of \aurora's heuristics below as examples, and refer readers to our artifact for additional examples~\cite{appendix}.

{\it Form screen}: Form screens usually contain multiple text fields, spinner components, and one submit button. Random-based tools often cannot go past these screens, as such screens require relevant text input. \aurora can provide the necessary knowledge for entering relevant input using its preset spreadsheet values. 
\safwatcrc{The values represent predefined column headers and associated data, which is static during application testing. 
We use a SentenceBERT model~\cite{reimers_sentence-bert_2019} to match input fields on UI screens with our preset spreadsheet values and input the top match to the UI screen. 
This model is essential for matching on-screen labels (e.g., ``last name'') with the relevant spreadsheet column names (e.g., ``surname''), which we then take a value from. 
The SentenceBERT model can resolve nuanced differences by capturing the semantic relationships between the labels and the spreadsheet column names.} 

{\it Player screen}: Player screens usually have a play and resume button and\Space{ a few} other smaller buttons\Space{ on screen}. AIG tools can get stuck in these screens, as the probability of hitting bigger buttons (e.g., play and resume) are higher, which does not necessarily result in a screen change. 
\aurora searches for other buttons, such as the settings or share button \safwatcrc{using its pre-defined heuristics and interacts with them. In this way, it helps}
with moving on to a different screen, so \AIGtools~can explore other parts of the application.

\section{Evaluation}
\label{sec:eval}

To understand how well AURORA can explore a given app, we ask the following research questions (RQs):

\begin{enumerate}

\item How well do \aurora's Screen Classifiers function in relation to a baseline?
\item How often do automated input generation tools get stuck, and what kind of screens are more difficult to explore?
\item How much improvement does \aurora offer over APE, Monkey, and VET regarding method coverage?
\item How often do the heuristics get executed successfully, and how many additional methods can they cover? 
\item How effective are the heuristics in navigating the intended tarpit screens?
\end{enumerate}

\subsection{Evaluation Context}

\begin{table}[t]
\centering
\footnotesize
\caption{Apps used for our evaluation. \#DLs represents the approximate number of downloads.}
\posttabcaptionspace
\label{appstats}
\begin{tabular}{llll}
\hline
App name       & Version  & Category               & \#DLs \\ \hline
\accuweather    & 7.4.1    & Weather                & 50m+  \\
\alltrails      & 14.2.0   & Travel \& Local        & 10m+  \\
\autoscout    & 9.8.0    & Auto \& Vehicles       & 10m+  \\
\carmax         & 2.56.1   & Auto \& Vehicles       & 5m+   \\
\duolingo       & 3.75.1   & Education              & 100m+ \\
\flipboard      & 4.1.1    & News \& Magazines      & 500m+ \\
\foxnews        & 4.50.0   & News \& Magazines      & 10m+  \\
\kayak          & 176.2    & Travel \& Local        & 10m+  \\
\merriamwebster & 4.1.2    & Books \& Reference     & 10m+  \\
\spotify        & 8.4.48   & Music \& Audio         & 100m+ \\
\tripadvisor    & 25.6.1   & Travel \& Local        & 100m+ \\
\trivago        & 4.9.4    & Travel \& Local        & 10m+  \\
\walmart        & 22.31    & Shopping               & 50m+  \\
\wattpad        & 6.82.0   & Books \& Reference     & 100m+ \\
\webtoon        & 2.4.3    & Comics                 & 10m+  \\
\wish           & 4.16.5   & Shopping               & 100m+ \\
\youtube        & 17.33.42 & Video Player \& Editor & 1b+   \\ \hline
\end{tabular}
\end{table}

\noindent\textbf{Datasets and Baselines:} We evaluate the two AURORA classifiers using the 1369 labeled RICO images derived as part of our UI Design Motif study. We compare our classifiers against Screen2Vec~\cite{li2021screen2vec}, which is a textual screen embedding technique that can be used to classify Android screens using a neural representation of UI metadata. \safwat{We evaluate the performance of APE, Monkey, VET, and \aurora on an emulator operating within the Android 6.0 environment. 
We run \aurora in conjunction with APE as the exploration tool due to its superior method coverage rate observed during the VET experimentation conducted by Wang et al. \cite{wang_vet_2021}. Additionally, APE has demonstrated a remarkable capacity to attain higher test coverage \cite{gu_practical_2019} compared to alternative tools, such as Monkey \cite{monkey_uiapplication_2022} or STOAT \cite{su_guided_2017}. Our experimental analysis focuses on a carefully selected set of 12 apps derived from the VET experiments. To ensure the validity of our findings, we exclude 4 apps from the previous study due to their lack of support on Android 6.0, which could potentially introduce inconsistencies in the results. Additionally, we expand our investigation to encompass five additional apps beyond the scope of the original VET experiments, resulting in a total of 17 apps under examination, one more than the number of apps assessed in the VET experiments~\cite{wang_vet_2021}.}

\safwat{The inclusion of the five additional apps is meant to help assess the generalizability of our heuristic-based approach. The additional five applications are \alltrails, \carmax, \foxnews, \kayak, and \walmart. These supplementary apps are among the most popular apps with over \Num{five} million downloads each. They are incorporated into the study to examine the broader applicability and generalizability of our proposed method.}
We also updated two apps (\autoscout\ and \youtube) to a newer version than the one used in the VET paper because almost all functionalities of those apps were disabled in the older versions at the time of our experiments.
Table~\ref{appstats} shows the statistics of the apps we used for our evaluation.

\newcommand{\DLs}{\#DLs}

\noindent{\textbf{Experimental Procedure: }}\safwat{Our experiment starts by executing Monkey and APE on 17 pre-selected apps.
We run a single instance of the emulator at a time to collect the app traces. The emulator is allocated 2 GB of RAM and 2 GB of internal storage space. To ensure the emulator remains responsive and efficient, we avoid running more than three 1-hour traces simultaneously.}
\safwatcrc{We conduct three 1-hour runs of Monkey and APE. VET learns from its built-in tarpit identification process from the APE runs and then adds three more 1-hour runs, giving us a total of six 1-hour runs. \aurora is executed for six 1-hour runs for each app. We compare the first three 1-hour runs of \aurora with APE and Monkey and compare the total six 1-hour runs with VET. Due to experiment costs, we do not conduct six 1-hour runs for Monkey and APE, therefore we present their comparison in a separate table. After each 1-hour run, our automated script restarts and wipes the emulator's data, preventing the emulator memory from filling up due to the screenshots taken during \aurora's runtime.}

\noindent\textbf{Metrics:} For \textbf{RQ$_1$}, we use the classic definitions of Precision, Recall, F1-score, and Accuracy for multi-class classification problems. 
For \textbf{RQ$_2$}, we consider a screen to be a tarpit if a tool gets stuck on the screen for more than 10 seconds, which we felt appropriate given the high number of actions tools like APE can generate.  
For \textbf{RQ$_3$}, we use MiniTrace~\cite{minitrace} to calculate the method coverage of our selected industrial apps. 
MiniTrace collects method coverage using the Android runtime and does not require app instrumentation. 
As MiniTrace requires Android 6.0 to run, we employ this Android version for our evaluation. 
We measure coverage by calculating the union of the method coverage over the set of three runs for each tool, which we refer to as ``set-union'' method coverage. We also calculate the area under the curve for our method coverage. This measurement tells us how soon a tool can achieve more coverage. The formula we use is
$\mathrm{AUC}=\sum_{i=1}^n \frac{1}{2}\left(R_{i-1}+R_i\right) \cdot \Delta t$, where $n$ is the total number test runs, $R_i$ is the method coverage at hour $i$, and $\Delta t$ represents the time interval for each one-hour test run.

\begin{table}[t]
\centering
\small
  \caption{Performance of \aurora{}'s Motif Classifiers.}
  \posttabcaptionspace
  \label{clipResults}
  \resizebox{0.4\textwidth}{!}{%
  \begin{tabular}{lrrrr}
    \toprule
    \textbf{} &\textbf{Precision} &\textbf{Recall} &\textbf{F1-Score} &\textbf{Accuracy}\\
    \midrule
        RICO &0.717 &0.686 &0.690 &0.689\\
        Extended &0.830 &0.812 &0.809 &0.813\\
  \bottomrule
\end{tabular}}
\end{table}

\begin{table}[t]
\centering
\small
\vspace{-0.5em}
\caption{Top 10 UI categories identified by \aurora in real-time and their associated bypassing rates. Bolded rows represent tarpit categories.} 
\posttabcaptionspace
\label{uiStopAurora}
\resizebox{0.45\textwidth}{!}{
\begin{tabular}{lrr}
\toprule
& \multicolumn{1}{l}{APE stopped} & \multicolumn{1}{l}{\% passed by \aurora} \\
\midrule
\textbf{Search screen}           & 633                                   & 93.7\%                    \\
Settings screen                  & 523                                   & 77.8\%                    \\
\textbf{Viewer screen}           & 479                                   & 97.1\%                    \\
Home menu screen                 & 473                                   & 78.9\%                    \\
\textbf{Onboarding screen}       & 444                                   & 90.3\%                    \\
Pop up menu                      & 418                                   & 99.5\%                    \\
\textbf{Web browser}             & 357                                   & 100\%                     \\
Catalog screen                   & 342                                   & 85.4\%                    \\
\textbf{Player screen}           & 312                                   & 81.7\%                    \\
\textbf{Log-in screen}           & 294                                   & 90.1\%                    \\
\bottomrule
Average & \apeStopAverage & \heurPassRate \\
\bottomrule
\end{tabular}}
\end{table}

\subsection{RQ1: Accuracy of Screen Classifiers?} 

As described in Section~\ref{approach:classifer}, \aurora was evaluated with Autoencoder-based and CLiP-based models. 
We find that\Space{ baseline classifier which follows an} \aurora{}'s Autoencoder-based model achieved $\approx$60\% accuracy, whereas Screen2Vec, our baseline, was able to achieve an overall accuracy of only $\approx$38\%. 
We get an even more sizeable increase in accuracy over our baseline with \aurora{}'s CLiP-based models. 
Table~\ref{clipResults} illustrates the classification effectiveness of \aurora's CLiP-based models on the 1369 labeled RICO images. 
The RICO variant is the performance without the unsupervised pre-training on the screens collected from Google, whereas the Extended model does include this process. 
Our results show that pre-training achieves\Space{has a measurable impact leading to} an 81.3\% accuracy compared to the other model's 68.9\% accuracy.

\subsection{RQ2: How often AIG Tools Get Stuck} 

Table~\ref{uiStopAurora} shows the top \Num{10} categories of UI screens considering the number of halts faced during the entire experimentation run. Search screens represent the most prevalent tarpit with 633 halts, but \aurora managed to navigate through \Num{93.7\%} of these screens using its heuristic-based approach. \aurora has an average of \Num{\heurPassRate} passing rate across all tarpit screens.

Previously, in our analysis of the VET dataset, we have identified certain categories that demonstrate characteristics akin to tarpits. These categories are denoted with bolded font in Table~\ref{uiStopAurora}. Notably, among the 8 UI classes previously identified as tarpit screens, 6 of them prominently feature within the top 10 UI screen categories.
This observation highlights the high potential for \aurora to improve \AIGtools{} as tarpit category screens frequently occur.

One tarpit category not in the top 10 is ``Advertisements'', which exhibited a relatively lower frequency in our experiments. This finding is reasonable, considering our experiment uses an older version of Android, that may no longer support certain ads for the applications~\cite{noauthor_why_nodate} we evaluated. 
Similarly, ``Form'' screens are not in the top 10 due to some apps no longer allowing sign-ups on older Android versions.

\subsection{RQ3: Method Coverage Improvement?} 
To evaluate the effectivness of \aurora, we compare the total number of unique methods covered over its three 1-hour runs to those of Monkey and APE. 
From Table~\ref{cumulCoverage}, we can see that \aurora gets an average of 41.4\% increase in coverage compared to Monkey. APE, on the other hand, gets a 28.2\% increase. If we combine a 1-hour APE run with 2-hour \aurora runs (denoted as APE1-AURORA2), we get the best performance, a 43.8\% increase from Monkey's method coverage. We also performed an experiment with two hours of APE runs combined with one hour of \aurora. However, the results were worse than just \aurora or the shown combination. The reason for the difference in improvement is likely due to the fact that APE1-AURORA2 best harnesses the strengths of each technique. That is, APE is able to exercise a large number of actions in a shorter period of time, whereas \aurora can more effectively explore tarpits, but benefits from the extra time budget to do so – due to its online classification and heuristic execution.

\revision{We can see that Monkey performs better than all of the other tools for the \duolingo{} app. In this app, the screens typically only require taps and the UI components cover large areas of the screen. 
As Monkey works by generating random events like taps or gestures without considering UI layouts/metadata, it has a higher action per second rate than APE or \aurora. 
While Monkey often suffers from empty space tap issues on other apps, it does not suffer this issue for \duolingo{} given its large components, and the high action rate leads to higher coverage.}
All in all, \aurora gets higher than APE in set union coverage for 13 out of 17 apps, while APE1-AURORA2 also gets higher coverage than APE for 13 apps. 

\begin{table}[t]

\footnotesize
\caption{Set union method coverage comparison. Bolded cells represent the highest coverage for an app across all tools.}
\posttabcaptionspace
\centering
\label{cumulCoverage}
\resizebox{0.5\textwidth}{!}{
\begin{tabular}{lrrrrrrr}
\hline
& \multicolumn{1}{l}{\textbf{Monkey}}   & \multicolumn{2}{l}{\textbf{APE}}                                   & \multicolumn{2}{l}{\textbf{\aurora}}                               & \multicolumn{2}{l}{\textbf{APE1-AURORA2}}                          \\ \cline{2-8} 
               \textbf{App}            & \multicolumn{1}{l}{Coverage} & \multicolumn{1}{l}{Coverage} & \multicolumn{1}{l}{\% inc} & \multicolumn{1}{l}{Coverage} & \multicolumn{1}{l}{\%inc} & \multicolumn{1}{l}{Coverage} & \multicolumn{1}{l}{\%inc} \\ \hline
\accuweather    & 16711                     & 21264                     & 27.2\%                     & 22641                     & 35.5\%                    & \textbf{22714}            & 35.9\%                    \\
\alltrails      & 28691                     & 43132                     & 50.3\%                     & \textbf{67231}            & 134.3\%                   & 59548                     & 107.5\%                   \\
\autoscout    & 29763                     & \textbf{40857}            & 37.3\%                     & 38554                     & 29.5\%                    & 39136                     & 31.5\%                    \\
\carmax         & 11002                     & 11619                     & 5.6\%                      & 17260                     & 56.9\%                    & \textbf{17452}            & 58.6\%                    \\
\duolingo       & \textbf{15328}            & 14355                     & -6.3\%                     & 14805                     & -3.4\%                    & 14993                    & -2.2\%                    \\
\flipboard      & 8652                      & 10646                     & 23.0\%                     & 13345                     & 54.2\%                    & \textbf{13569}            & 56.8\%                    \\
\foxnews        & 27705                     & 29924                     & 8.0\%                      & 30574                     & 10.4\%                    & \textbf{31375}            & 13.2\%                    \\
\kayak          & 44593                     & 55688                     & 24.9\%                     & 57555                     & 29.1\%                    & \textbf{59327}            & 33.0\%                    \\
\merriamwebster & 7668                      & 8621                      & 12.4\%                     & 9112                     & 18.8\%                    & \textbf{9175}             & 19.7\%                    \\
\spotify        & 12510                     & 19533                     & 56.1\%                     & \textbf{28552}            & 128.2\%                   & 27071                     & 116.4\%                   \\
\tripadvisor    & 23390                     & \textbf{30548}            & 30.6\%                     & 27728                     & 18.5\%                    & 30047                     & 28.5\%                    \\
\trivago        & 19296                     & 20096                     & 4.1\%                      & \textbf{20393}            & 5.7\%                     & 20343                     & 5.4\%                     \\
\walmart        & 27322                        & 40435                        & 48.0\%                     & 44041                        & 61.2\%                    & \textbf{51149}               & 87.2\%                    \\
\wattpad        & 13324                        & 23426                        & 75.8\%                     & 23648                        & 77.5\%                    & \textbf{24690}               & 85.3\%                    \\
\webtoon        & 19310                        & 27628                        & 43.1\%                     & 22819                        & 18.2\%                    & \textbf{27750}               & 43.7\%                    \\
\wish           & 7544                         & 9175                         & 21.6\%                     & \textbf{9192}                & 21.8\%                    & 8450                         & 12.0\%                    \\
\youtube        & 32428                        & \textbf{38372}               & 18.3\%                     & 34738                        & 7.1\%                     & 36030                        & 11.1\%                    \\ \hline
Average        & \multicolumn{1}{l}{}         & \multicolumn{1}{l}{}         & \textbf{28.2\%}            & \multicolumn{1}{l}{}         & \textbf{41.4\%}           & \multicolumn{1}{l}{}         & \textbf{43.8\%}           \\ \hline
\end{tabular}}
\end{table}

Considering \aurora vs APE1-AURORA2, we see that the latter is clearly ahead in set union method coverage and area under the curve. This result indicates that, for a 3-hour run, \aurora should be combined with APE to get the best possible coverage.

To run VET, we must first run three hours of APE, and then, learning from the actions that end up in a stuck region, VET prevents them from happening in its additional 3-hour run. 
To make a fair comparison, we run \aurora for 6 hours. 
Table~\ref{vetVSaurora} compares the set union methods and exclusive methods for VET and \aurora. 
Considering coverage, \aurora gets an average of \percImprovVET increase compared to VET. 
We can also see that \aurora gets higher coverage for 16 out of 17 apps. Considering orthogonality, \aurora provides an average of 19.6\% exclusive methods compared to 6.3\% from VET.

\begin{table}[t]
\caption{Set union method coverage of VET vs. \aurora. Bolded cells represent the highest coverage between the tools.}
\vspace{-1ex}
\label{vetVSaurora}
\centering
\resizebox{0.5\textwidth}{!}{
\begin{tabular}{lrrrrrrrr}
\hline
\textbf{App}     & \multicolumn{1}{c}{\textbf{VET}} & \multicolumn{1}{c}{\textbf{AU}} & \multicolumn{1}{c}{\textbf{\% inc}} & \multicolumn{1}{c}{\textbf{Comm.}} & \multicolumn{1}{c}{\textbf{V ex (\%)}} & \multicolumn{1}{c}{\textbf{AU ex (\%)}} \\ \hline
\accuweather      & 23456                                     & \textbf{28929}                               & 23.3\%                                       & 22105                                       & 1351 (\z{}4.5\%)                                      & 6824 (22.5\%) \\
alltrails        & 43765                                     & \textbf{68829}                               & 57.3\%                                       & 43256                                       & 509 (\z{}0.7\%)                                        & 25573 (36.9\%) \\
\autoscout      & 41258                                     & \textbf{42953}                               & 4.1\%                                        & 38478                                       & 2780 (\z{}6.1\%)                                      & 4475 (\z{}9.8\%) \\
\carmax           & 12331                                     & \textbf{19725}                               & 60.0\%                                       & 11876                                       & 455 (\z{}2.3\%)                                       & 7849 (38.9\%) \\
\duolingo         & 14704                                     & \textbf{15628}         & 6.3\%                                        & 14291                                       & 413 (\z{}2.6\%)                                       & 1337 (\z{}8.3\%)  \\
\flipboard        & 11754                                     & \textbf{14705}                               & 25.1\%                                       & 10830                                       & 924 (\z{}5.9\%)                                       & 3875 (24.8\%)  \\
\foxnews          & 31140                                     & \textbf{31586}                               & 1.4\%                                        & 29601                                       & 1539 (\z{}4.6\%)                                       & 1985 (\z{}6.0\%)  \\
\kayak            & 57641                                     & \textbf{77567}                               & 34.6\%                                       & 55103                                       & 2538 (\z{}3.2\%)                                       & 22464 (28.0\%) \\
Merriam-Web   & \textbf{10547}                            & 9734                                         & -7.7\%                                       & 9328                                        & 1219 (11.1\%)                                       & 406 (\z{}3.7\%)  \\
\spotify          & 19918                                     & \textbf{32111}                               & 61.2\%                                       & 19555                                       & 363 (\z{}1.1\%)                                       & 12556  (38.7\%) \\
\tripadvisor      & 32014                                     & \textbf{32200}                               & 0.6\%                                        & 29973                                       & 2041 (\z{}6.0\%)                                      & 2227 (\z{}6.5\%) \\
\trivago          & 20265                                     & \textbf{20944}                               & 3.4\%                                        & 20032                                       & 233 (\z{}1.1\%)                                       & 912 (\z{}4.3\%) \\
\walmart          & 43334                                     & \textbf{46232}                               & 6.7\%                                        & 35194                                       & 8140 (15.0\%)                                       & 11038 (20.3\%) \\
\wattpad          & 24053                                     & \textbf{33192}                               & 38.0\%                                       & 23176                                       & 877 (\z{}2.6\%)                                       & 10016 (29.4\%) \\
\webtoon          & 28059                                     & \textbf{31789}                               & 13.3\%                                       & 20940                                       & 7119 (18.3\%)                                      & 10849 (27.9\%) \\
\wish             & 9923                                      & \textbf{10305}                               & 3.8\%                                        & 8178                                        & 1745 (14.5\%)                                       & 2127 (17.7\%) \\
\youtube          & 41518                                     & \textbf{42466}          & 2.3\%                                        & 37993                                       & 3525 (\z{}7.7\%)                                       & 4473 (\z{}9.7\%) \\ \hline
\textbf{Average} & \multicolumn{1}{l}{}                      & {}                        & \textbf{19.6\%}                              & 25288.8                                     & 2104.2 (\textbf{\z6.3\%})                                     & 7587.4 (\textbf{19.6\%})  \\ \hline
\end{tabular}}
\end{table}

\begin{table}[t]
\footnotesize
\vspace{-1em}
\caption{Heuristics success rate across all apps.}
\posttabcaptionspace
\centering
\label{heurSuccessApp}
\begin{tabular}{lrrrr}
\hline
\textbf{App}     & \multicolumn{1}{l}{\textbf{Passed}} & \multicolumn{1}{l}{\textbf{Failed}} & \multicolumn{1}{l}{\textbf{Total}} & \multicolumn{1}{l}{\textbf{\% Pass}} \\ \hline
\accuweather      & 421                                 & 144                                 & 565                                & 74.5\%                               \\
\alltrails        & 308                                 & 35                                  & 343                                & 89.8\%                               \\
\autoscout      & 420                                 & 36                                  & 456                                & 92.1\%                               \\
\carmax           & 275                                 & 18                                  & 293                                & 93.9\%                               \\
\wish             & 242                                 & 21                                  & 263                                & 92.0\%                               \\
\duolingo         & 364                                 & 31                                  & 395                                & 92.2\%                               \\
\foxnews          & 279                                 & 12                                  & 291                                & 95.9\%                               \\
\youtube          & 532                                 & 5                                  & 537                                & 99.1\%                               \\
\kayak            & 251                                 & 39                                  & 290                                & 86.6\%                               \\
\merriamwebster   & 438                                 & 31                                  & 469                                & 93.4\%                               \\
\webtoon          & 219                                 & 68                                  & 287                                & 76.3\%                               \\
\spotify          & 278                                 & 40                                  & 318                                & 87.4\%                               \\
\tripadvisor      & 177                                 & 49                                  & 226                                & 78.3\%                               \\
\trivago          & 330                                 & 53                                  & 383                                & 86.2\%                               \\
\walmart          & 320                                 & 32                                  & 352                                & 90.9\%                               \\
\flipboard        & 346                                 & 49                                  & 395                                & 87.6\%                               \\
\wattpad          & 301                                 & 30                                  & 331                                & 90.9\%                               \\ \hline
\textbf{Total} & 5501                               & 693                                & 6194                                & \textbf{\heurPassRate}                      \\ \hline
\end{tabular}
\end{table}

\begin{figure}[t]
\centering
  \includegraphics[width=0.462\textwidth]{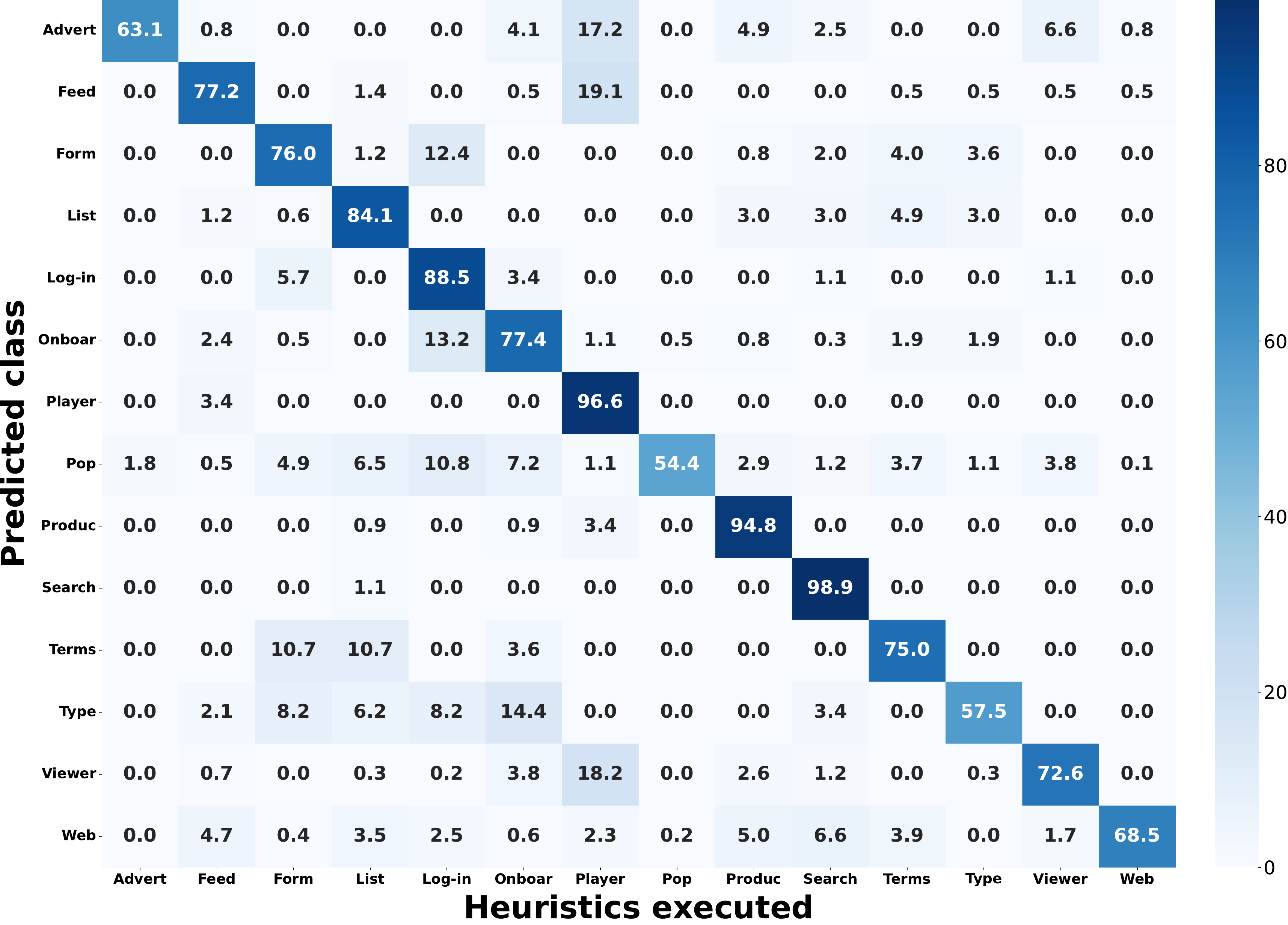}
  \vspace{-1ex}
  \caption{Confusion matrix of heuristics succeeding.}
  \label{fig:confusion_matrix}
\end{figure}

\subsection{RQ4: Successful Heuristic Execution?}
Table~\ref{heurSuccessApp} shows the success rate of our heuristics\Space{ among all apps}. 
Ranging from 74.5\% to 99.1\% with an average of \heurPassRate of the executed heuristics being successful. We consider our heuristics successful when any of the heuristics from \aurora's top 3 predictions are successful in\Space{ making} changing the app screen.\Space{ This tells us that the heuristics made some changes to the screen, whereas a random-based tool was stuck for 10 seconds.}

If we compare actions per second, using \aurora will generate fewer actions than not using \aurora, as it needs to classify screens and run heuristics during runtime. 
However, even with less actions generated, \aurora still achieves an improvement in code coverage. 
This result suggests that at tarpit screens,\Space{ where random-based tools get stuck,} a properly curated heuristic is often better than randomly clicking around to increase code coverage.

\revision{\aurora would likely offer similar improvements to systematic testing approaches, as they may also get “stuck” on screens\Space{ as a limitation of the systematic input generation scheme}. However, the VET dataset, and identified tarpits, were derived using random-based techniques, thus we oriented our analysis toward these techniques as well. Future work should explore \aurora’s effect on other types of techniques.}

\subsection{RQ5: Effectiveness of Individual Heuristics}
The \heurPassRate success rate of heuristics shows us the collective effectiveness of using a heuristic-based approach. To understand the performance of each of these heuristics to effectively navigate their respective UI tarpit screens, we conducted an analysis using \aurora's post-run logs. 

Our heuristic based approach works by iterating through the top three predictions of a tarpit screen. \aurora runs the heuristic designed for the initial predicted UI category, and when it does not result in a change in the tarpit screen, it proceeds to the subsequent predictions in a sequential manner. Figure~\ref{fig:confusion_matrix} shows the frequency with which various heuristics successfully made changes to a predicted screen, presented in a percentage format. The diagonal values refer to the heuristic affecting changes being the one intended for the first prediction. 
As the figure suggests, we find that \aurora was able to navigate all the predicted screens using their intended heuristics most of the time. Outside the diagonals, we can see that player heuristics also successfully navigated a handful of advertisement, feed, and viewer screens. 
This result is due to player heuristic's ability to find\Space{ different} on-screen components and resume random exploration on a different app screen.

\section{Threats to Validity}
\label{sec:threats}

Our initial study on design motifs involved manual effort in classifying screens, identifying UI tarpits, and finding ways to overcome them. Any manual process can include biases. We limit the potential for bias by examining only a portion of tarpit screens during heuristic construction and using semantic text matching to make our heuristics generalizable.

Another threat to our work's external validity is that we use only Android 6.0\Space{ for our testing platform, which is fairly old}. We utilize the MiniTrace mechanism for collecting method coverage without needing code instrumentation, and MiniTrace works with only Android 6.0. 
\section{Related Work}
\label{sec:related-work}

\noindent\textbf{Studies on Android Testing:} V{\'{a}}squez et al.~\cite{linares-vasquez_mining_2018} compiled a body of knowledge that can help researchers focus on new automated testing approaches tailored to developer needs. Choudhary et al.~\cite{choudhary_automated_2015} analyzed various modern test generation tools in a systematic way and illustrated that, surprisingly, the simpler Monkey tool surpassed more sophisticated tools in terms of code coverage, ease of use, and fault detection.

\noindent\textbf{Random-based Testing Tools:} Random based testing approaches construct test cases in a pseudo-random manner from the set of all possible program inputs~\cite{orso_software_2014}. 
Random-based tools excel in adaptability, as they target the app under test only on a per-screen basis. This technique was popularized in the Android testing tool, Monkey~\cite{monkey_uiapplication_2022}, and was later adapted by APE~\cite{gu_practical_2019}, Dynodroid~\cite{machiry_dynodroid_2013}, Intent Fuzzer~\cite{sasnauskas_intent_2014}, and VANARSena~\cite{ravindranath_automatic_2014}.

\noindent\textbf{Model-based Testing Tools:} MonkeyLab~\cite{moran_automatically_2016} uses the GUI-based models extracted from Android application execution traces to generate usage scenarios.
The results demonstrate that MonkeyLab is able to generate effective and fully replayable scenarios. 
Moran et al.~\cite{moran_automatically_2016} studied the importance of crashes during Android application testing. The authors developed CrashScope, a tool that can automatically discover, report, and reproduce crashes. They executed their tool on 61 Android apps and compared their tool with A3E, DynoDroid, MobiGUITAR, Monkey, and Puma~\cite{machiry_dynodroid_2013,hao_puma_2014,azim_targeted_2013,amalfitano_mobiguitar_2015,monkey_uiapplication_2022}.

Dong et al.~\cite{dong_time-travel_2020} proposed time-travel testing for Android, which works to maximize exploration efficiency by resuming to the most progressive states observed in the past. They evaluated their approach against Sapienz~\cite{mao_sapienz_2016} and Stoat~\cite{su_guided_2017} and it outperformed them in\Space{ statement} coverage and crashes discovered. 

\noindent\textbf{Machine Learning-based Testing Tools:} Li et al.~\cite{li_humanoid_2019} propose Humanoid, a testing tool that uses a combination of CNN and Residual LSTM in their approach to generate automated tests. They use the RICO dataset~\cite{deka_rico_2017} and perform CNN on the screenshots to predict action location on a given UI screen. Residual LSTM is used to predict the type of action performed - such as tap, long tap, swipe, etc. Q-testing~\cite{pan_reinforcement_2020} is an AIG tool that uses reinforcement learning for input generation. 

QTypist~\cite{liu_fill_2022} is a tool designed to automate the generation of input text for mobile applications\Space{. QTypist does so} by interacting with a large language model. 
Compared to \aurora, which handles various different UI exploration challenges, QTypist focuses on only text-related challenges.\Space{ A prompt-based approach, like QTypist, can take a lot of time to generate text, which can be expensive when trying to achieve high coverage in a short amount of time\Space{in a time-constrained test coverage task}.}
\section{Conclusion}
\label{sec:conclusions}

In this paper, we proposed AURORA, a framework that runs alongside \AIGtools{} and can categorize and navigate around UI screens when an \AIGtool{} is stuck using multimodal techniques for neural screen understanding.
Our evaluation illustrates that \aurora can effectively recognize different types of screens and effectively navigate around them, increasing the effectiveness of \AIGtools{}.
To aid future research, we make \aurora\Space{ including the classifiers used in our evaluation} and our labeled dataset of UI categories publicly available~\cite{appendix}.

\IEEEpeerreviewmaketitle

\balance
\pagebreak
\bibliographystyle{IEEEtran}
\bibliography{references2}

\end{document}